\documentstyle[twoside]{article}
\newcount\Mac  \Mac=0  
\newcommand{\ifMac}[2]{\ifnum\Mac=1 #1 \else #2 \fi}
\def\putps(#1,#2)(#3,#4)#5#6{\ifnum\Mac=1 \put(#1,#2){\special{picture #5}}
\else  \put(#3,#4){\includegraphics{#6}} \fi}
\def\Red  {}
\def\Black{}
\def\Blue {}
\newcommand{\GeV}{\,{\rm GeV}}
\newcommand{\eV}{\,{\rm eV}}

\newcommand{\MeV}{\,{\rm MeV}}
\newcommand{\NP}{Nucl. Phys.}
\newcommand{\Jhep}{{\sc J.hep}}

\newcommand{\PL}{Phys. Lett.}
\newcommand{\PR}{Phys. Rev.}

\newcommand{\eq}[1]{~(\ref{eq:#1})}
\newcommand\degree{{}^{\circ}}

\def\circa#1{\,\raise.3ex\hbox{$#1$\kern-.75em\lower1ex\hbox{$\sim$}}\,}
\makeatletter
\def\art{\@ifnextchar[{\eart}{\oart}}
\def\eart[#1]#2#3#4#5#6{{\rm #2}, {\em #3 \rm #4} {\rm (#6) #5 ({\em #1})}}
\def\hepart[#1]#2{{\rm #2, \em#1}}
\newcommand{\oart}[5]{{\rm #1}, {\em #2 \rm #3} {\rm (#5) #4}}

\newcounter{alphaequation}[equation]
\def\thealphaequation{\theequation\hbox to
0.6em{\hfil\alph{alphaequation}\hfil}}
\def\eqnsystem#1{
\def\@eqnnum{{\rm (\thealphaequation)}}
\def\@@eqncr{\let\@tempa\relax \ifcase\@eqcnt \def\@tempa{& & &} \or
  \def\@tempa{& &}\or \def\@tempa{&}\fi\@tempa
  \if@eqnsw\@eqnnum\refstepcounter{alphaequation}\fi
\global\@eqnswtrue\global\@eqcnt=0\cr}
\refstepcounter{equation} \let\@currentlabel\theequation \def\@tempb{#1}
\ifx\@tempb\empty\else\label{#1}\fi
\refstepcounter{alphaequation}
\let\@currentlabel\thealphaequation
\global\@eqnswtrue\global\@eqcnt=0 \tabskip\@centering\let\\=\@eqncr
$$\halign to \displaywidth\bgroup \@eqnsel\hskip\@centering
$\displaystyle\tabskip\z@{##}$&\global\@eqcnt\@ne
\hskip2\arraycolsep\hfil${##}$\hfil& \global\@eqcnt\tw@\hskip2\arraycolsep
$\displaystyle\tabskip\z@{##}$\hfil
\tabskip\@centering&\llap{##}\tabskip\z@\cr}
\def\endeqnsystem{\@@eqncr\egroup$$\global\@ignoretrue} \makeatother

\begin{document}
\twocolumn[
\centerline{2 Apr.\ 1999 \hfill    IFUP--TH/14--99}
\centerline{hep-ph/9904245 \hfill } \vspace{1cm}
\centerline{\LARGE\bf\Red Oscillations of three neutrinos with all $\Delta m^2\sim 10^{-3}\eV^2$}

\bigskip\bigskip\Black
\centerline{\large\bf Alessandro Strumia} \vspace{0.3cm}

\centerline{\em Dipartimento di Fisica, Universit\`a di Pisa and}
\centerline{\em INFN, sezione di Pisa,  I-56126 Pisa, Italia}\vspace{0.3cm}

\bigskip\bigskip\Blue

\centerline{\large\bf Abstract}
\begin{quote}\large\indent
Oscillations of three neutrinos with all squared mass splittings around $10^{-3}\eV^2$
are not firmly excluded by solar neutrino experiments. We carefully
verify that they are also perfectly compatible
with atmospheric neutrino experiments:
due to accidental reasons the SuperKamiokande experiment is rather insensitive to `solar' $\nu_e/\nu_\mu$ oscillations,
even if some characteristic small effects could become visible with more statistics.
This pattern of oscillations can be 
excluded by new solar experiments, or cleanly discovered at {\sc KamLand}.

We also perform a fit of the most recent atmospheric SK data
within the usual assumption that `solar' effects are negligible.
\end{quote}\Black
\vspace{1cm}]

\noindent

\section{Introduction}
In this paper we explore the possibility that the atmospheric and solar neutrino deficits 
can be produced by oscillations of the three known neutrinos with
comparable mass splittings $\Delta m^2\sim 10^{-3}\eV^2$ and
large $\nu_e/\nu_\mu$ and $\nu_\mu/\nu_\tau$ mixings.
In section~\ref{sole} we recall why solar neutrino experiments do not exclude this possibility.
In section~\ref{atm+sole} we show that this possibility is also perfectly consistent with atmospheric neutrino experiments,
even if it is sometimes said that, since SuperKamiokande (SK) sees no
anomaly in the rate of $\nu_e$ events,
significant oscillations of atmospheric $\nu_e$ neutrinos are excluded.
This is not the case~\cite{lungo}. In short the reason is the following:
atmospheric neutrinos are produced by cosmic rays in the following proportion
$$(N_{\nu_e},N_{\nu_\mu},N_{\nu_\tau})\propto(1,R,0).$$
Since $R\approx 2$ the nearly maximal $\nu_\mu/\nu_\tau$ oscillation responsible of the atmospheric $\nu$ anomaly
gives oscillated neutrinos with composition $\propto(1,1,1)$ ---
the only proportion not affected by further possible oscillations.
Of course this argument is only approximate
($R$ is larger than 2 for $E_\nu\circa{>}1\GeV$ and the $\nu_\mu/\nu_\tau$ oscillation
need not be exactly maximal): a numerical computation
will confirm that the conclusion is correct.

\smallskip

From a theoretical point of view, the possibility of explaining neutrino anomalies
with comparable $\Delta m^2$ has important consequences.
It is easy to build models that naturally explain large mixings, but between neutrinos with comparable mass.
It is also easy to build models that give hierarchical neutrinos, but with small mixings.
It is more difficult to obtain large mixing angles ($\theta_{23}\sim 1$) between
hierarchical neutrinos ($\Delta m^2_{23}\gg \Delta m^2_{12}$,
all oscillation parameters are precisely defined later on in eq.\eq{V}): only
few mass matrices (justifiable with various symmetries) naturally give this pattern~\cite{lungo,textures}.
Thus, non hierarchical neutrinos would not give very restrictive indications on flavour physics.

\medskip

The paper is structured as follows.
In section~\ref{sole} we explain why a large solar $\Delta m^2\sim10^{-3}\eV^2$
is not safely excluded by solar neutrino experiments.
In section~\ref{fit} we describe how we will fit the SK data.
Since this is a delicate task, in section~\ref{atm-fit} we test our procedure
performing a complete fit of the most recent SK data, in the `standard' case where the solar $\Delta m^2_{12}$
is negligibly small.
In section~\ref{atm+sole} we show that SK data are perfectly compatible with
a $\Delta m^2_{12}$ as large as allowed by the {\sc Chooz} bound~\cite{CHOOZ},
and we discuss how its effects can be detected at SK and at future experiments.

\section{Energy independent solar oscillations?}\label{sole}
If solar and atmospheric neutrino experiments are not
affected by unknown systematic errors or by rare statistical fluctuations,
if the standard solar models (SSMs) are correct,
if there are only the three known light neutrinos, then
forthcoming neutrino experiments will
confirm and measure more precisely that
\begin{equation}\begin{array}{ll}
\Delta m^2_{23}\approx 10^{-3}\eV^2 & \sin^2 2\theta_{23}\approx 1\\
\Delta m^2_{12}\approx 10^{-10}\eV^2 & \sin^2 2\theta_{12}\approx 1\\
\end{array}\end{equation}
These values give a good fit of atmospheric neutrino data and give an
acceptable fit ($10\%$ C.L.~\cite{BKS,noMSW,9903262}) of solar neutrino data.

\medskip

Since these conclusions are quite strong, it is useful to discuss if they are also strongly founded.

\smallskip

The well known `MSW solutions' with $\Delta m^2\approx 10^{-5}\eV^2$ give a poor fit of
the distortion of the solar $^8$B spectrum observed by SuperKamiokande~\cite{noMSW,9903262}
and are ruled out at 95\% C.L.~\cite{noMSW}.
Maybe this experimental result (or the estimation of its uncertainties) is wrong.
Maybe the distortion is not produced by $\nu$ oscillations,
but is due to a flux of `hep' neutrinos $\sim15$ times
higher than what predicted by SSMs~\cite{lungo,hep}.


At the moment, it seems more safe to believe that there are at least three (instead of one)
possible oscillation solutions to the solar neutrino problem.
We now discuss why even this sentence is not strongly founded:
it is not safely excluded that the solar neutrino anomaly can be explained by
an energy independent $\nu_e\to \nu_e$ survival probability $P_{ee}\sim 1/2$
(as can be produced by a large $\Delta m^2_{12}\sim 10^{-3}\eV^2$).

\smallskip

The deficit $r_i=\Phi_i^{\rm exp}/\Phi_i^{\rm BP98}$
of solar neutrinos measured by the three kind of solar experiments
($i=$ Cl, Ga and SK),
with respect to the central values $\Phi_i^{\rm BP98}$ of fluxes predicted by the BP98~\cite{BP98} SSM, are
\begin{eqnsystem}{sys:sunexp}
r_{\rm Cl}&=&0.315\pm0.025~~\cite{ClSun}\\
r_{\rm SK}&=&0.47\pm0.02~~~~~\cite{KaSun}\\
r_{\rm Ga}&=&0.58\pm0.05~~~~~\cite{GaSun}
\end{eqnsystem}
(the errors do not include the SSM uncertainty).
The predictions of a solar-model-independent analysis, in presence of an energy-independent $P_{ee}$, are
\begin{eqnsystem}{sys:sunth}
r_{\rm Cl}&=& P_{ee}(0.03+0.73 R_{\rm {}^8B}+0.24 R_{\rm {}^7Be})\qquad\\
r_{\rm SK}&=& (0.15 +0.85 P_{ee}) R_{\rm {}^8B}\\
r_{\rm Ga}&=& P_{ee}(0.60+0.10 R_{\rm {}^7Be}+0.31 R_{\rm {}^8B})
\end{eqnsystem}
where $R_\alpha\equiv \Phi_\alpha/\Phi_\alpha^{\rm BP98}$ is the ratio of total flux of type $\alpha$ neutrinos
($\alpha = \rm pp, p\hbox{$e$}p, ^7\!Be,^{13}\!N,^{15}\!O,{}^{17}\!F,{}^8\!B,\break h\hbox{$e$}p$) emitted
by the sun, with respect to the central value of the BP98~\cite{BP98} SSM.
Only two free parameters $ R_{\rm {}^8B}$ and $R_{\rm {}^7Be}$ appear in eq.s~(\ref{sys:sunth});
the others have been eliminated using the fact that the total luminosity of the sun is known,
and other solid informations (see~\cite{CDFLR,lungo} for more details).

\begin{figure*}[t]
\begin{center}
\begin{picture}(17.7,5)
\putps(-0.5,0)(-0.5,0){fitchiq}{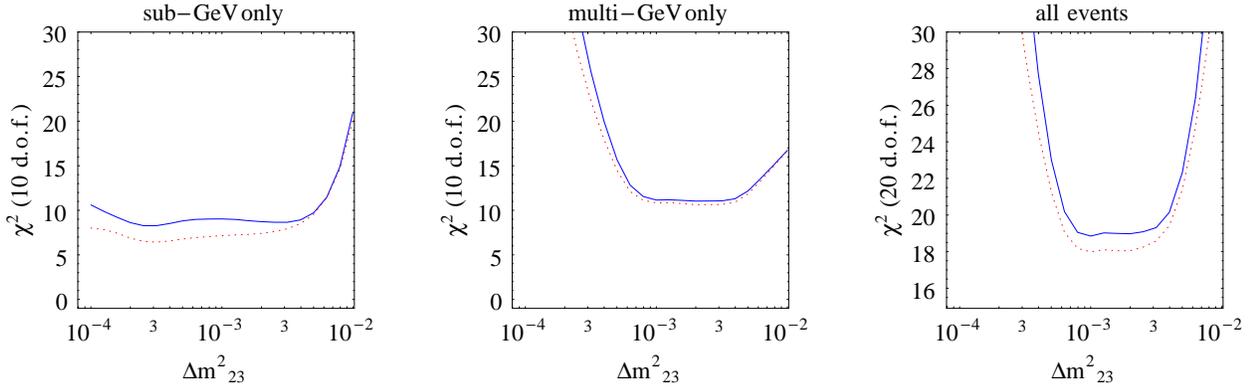}
\end{picture}
\caption[SP]{\em $\chi^2$ of SuperKamiokande data as function of
$\Delta m^2_{23}$ for $\theta_{23}=45\degree$, $\theta_{13}=0$ and negligible $\Delta m^2_{12}$.
In fig.~{\rm\ref{fig:chiq}a (b)} we show the separate contribution from sub-GeV (multi-GeV) events;
while all events are used in fig.~{\rm\ref{fig:chiq}c}.
Continuous and dotted lines correspond to two different definitions of the $\chi^2$ (see the text).
\label{fig:chiq}}
\end{center}\end{figure*}

\noindent
From eq.s~(\ref{sys:sunexp}) and~(\ref{sys:sunth}) we can easily see that
\begin{itemize}

\item If BP98 is correct (it predicts $R_{\rm {}^8B}=1\pm 0.2$ and $R_{\rm {}^7Be}=1\pm0.05$~\cite{BP98}),
the three experimental data are compatible with
an energy-independent $P_{ee}$
only in presence of a very unprobable statistical fluctuation ($p\approx 0.2\%$)~\cite{PeeCteBad,lungo};

\item Even if SSMs are not correct
(i.e.\ $R_{\rm {}^8B}$ and $R_{\rm {}^7Be}$ are treated as free parameters of order one),
the experimental data are incompatible with 
an energy-independent $P_{ee}$~\cite{PeeCteBad,lungo}.
A decent fit is possible only if SSMs are so wrong
($R_{\rm {}^7Be}$ close to zero) that the solar neutrino problem disappears~\cite{lungo} ---
a possibility strongly disfavoured by recent helio-sysmological tests of SSMs.

\item Solar data can be explained by
an energy-inde\-pen\-dent $P_{ee}$ if BP98 is correct, but one of the experiments is
affected by some unknown systematic error~\cite{lungo}.
For example, {\em  
it is sufficient to double the error quoted by the chlorine experiment~\cite{ClSun}
to have $P_{ee}=1/2$ compatible with experimental results}
(in presence of a reasonably probable, $p\sim 10\%$, statistical fluctuation).

\end{itemize}
In conclusion we believe that the evidence for a deficit of solar neutrinos is strong,
while there is yet no strong evidence for an energy dependent oscillation of solar neutrinos.

\section{Fitting SK data}\label{fit}
Reproducing what SK has really measured is a complex and delicate task.
We briefly describe how we do the computation.
Five basic ingredients are necessary for a fit of the SK data.

\begin{enumerate}

\item The experimental data: we use the most recent ones (736 days of data taking)~\cite{SKexp}.

\item The prediction for the flux of atmospheric neutrinos produced by cosmic rays~\cite{Gaisser}.
We include effects due to the magnetic field of the earth and to variation of solar activity.

\item The oscillation probability for neutrinos across the earth and the atmosphere.
In our case we have a generic neutrino $3\times 3$ mass matrix with 3 comparable $\Delta m^2$:
since we know no simple analytic approximation that takes into account all potentially relevant matter effects
(the MSW~\cite{MSW} effect and resonances that can affect the neutrinos that cross the 
mantle and the core of the earth~\cite{ParRes})
we include all matter effects with a fully numerical computation.
The disadvantage is that the transition probabilities for oscillations
with long pathlength $L\gg E_\nu /\Delta m^2$ are rapidly oscillating functions of the neutrino energy.
Even for the simplest realistic model of earth density, a numerical averaging
(using a sufficiently large number of $E_\nu$-bins) is more efficient than
analytic averaging.

\item The cross section and detection efficiencies in the SK detector.
A simple and safe technique has been used in~\cite{SKfit} for fitting the old Kamio\-kande data.
They employed the energy spectra of parent neutrinos given by
the Monte Carlo simulation of the Kamiokande detector.
The corresponding data for the much larger SK detector have not been published 
(without these information it is not possible to know what SK is really measuring),
although they are now available at the www address~\cite{www}.
We take into account that SK measures the neutrino direction with an
error that depends on the $\nu$ energy~\cite{tesi}
($\delta \theta$ is around $60\degree$ in sub-GeV events and $17\degree$ in multi-GeV events).

~~We do not include in the fit data about `upward through going muons'~\cite{mu-roccie}
because they are subject to larger theoretical uncertainties and
they are too energetic for being strongly affected by oscillations
(they are however very interesting for excluding alternative 
explanations of the atmospheric $\nu$ deficit~\cite{osc!}).

\item A $\chi^2$ function.
This is a delicate point, since it requires an estimation of
theoretical uncertainties in neutrino fluxes and correlation between them.
For simplicity we stick to the accurate definition of~\cite{fogli}
and we will discuss when appropriate the effect of different definitions.

\end{enumerate}
Moreover we impose the {\sc Chooz} bound about disappearance of reactor $\bar{\nu}_e$~\cite{CHOOZ}.
The {\sc Chooz} collaboration measures the annihilation energy
$E=E_{{\bar \nu}_e}+m_{\rm p}-m_{\rm n}+(2-1)m_e$
of positrons produced by
inverse beta decay $\bar{\nu}_e {\rm p}\to \bar{e}{\rm n}$
for various energy bins between $1$ and $7\MeV$.
Since we will be interested in an oscillation pattern where the $\bar{\nu}_e$ survival probability
depends on $E_{\bar{\nu}_e}$, we carefully treat the {\sc Chooz} data,
grouping them into two $E_{\bar{\nu}_e}$ bins.
We correctly reproduce the `initial' {\sc Chooz} bound ($\Delta m^2_{12}<0.9~10^{-3}\eV^2$
for maximal mixing at $90\%$ C.L.).
A `final' analysis of the whole {\sc Chooz} data has not yet been presented:
with increased statistics, the {\sc Chooz} bound could be improved up to $\approx0.6~10^{-3}\eV^2$~\cite{CHOOZ2}.

\begin{figure*}[t]
\begin{center}
\begin{picture}(17.7,5)
\putps(-0.5,0)(-0.5,0){fitstandard}{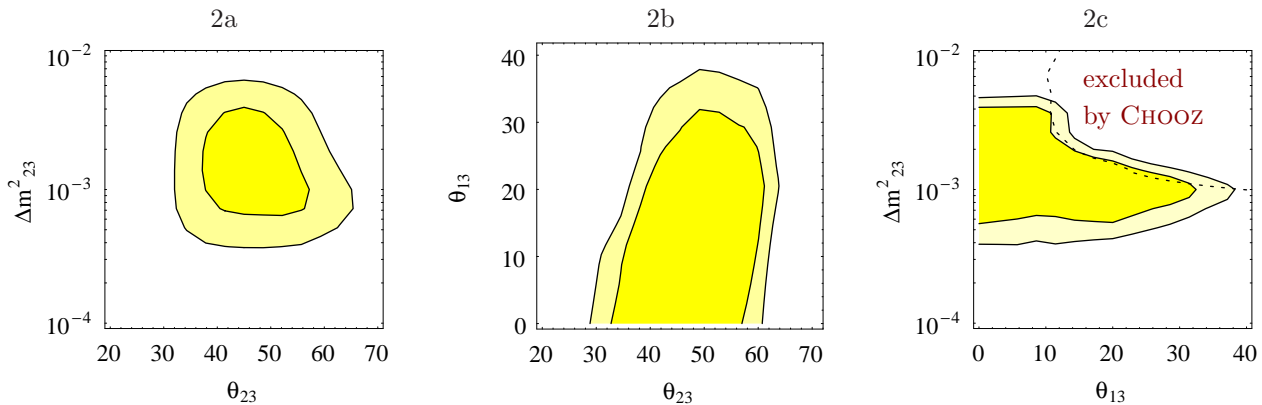}
\put(2.7,5.3){\ref{fig:fitstandard}a}
\put(8.5,5.3){\ref{fig:fitstandard}b}
\put(14.3,5.3){\ref{fig:fitstandard}c}\Red
\put(14.3,4.5){excluded}
\put(14.3,4){by {\sc Chooz}}\Black
\end{picture}
\caption[SP]{\em Fit of SK and {\sc Chooz} data under the assumption that `solar' oscillations are negligible,
so that the only relevant oscillations parameters are $\Delta m^2_{23}$, $\theta_{23}$ and $\theta_{13}$.
Inside the darker (lighter) areas $\chi^2<24~(30)$
(the $\chi^2$ uses $22$ experimental data; the best fit has $\chi^2_{\rm min}=18$;
In fig.~\hbox{\rm\ref{fig:fitstandard}a} the $\chi^2$ is minimized with respect to $\theta_{13}$,
in fig.~\hbox{\rm\ref{fig:fitstandard}b} with respect to $\Delta m^2_{23}$, and 
in fig.~\hbox{\rm\ref{fig:fitstandard}c} with respect to $\theta_{23}$.
\label{fig:fitstandard}}
\end{center}\end{figure*}

\section{Standard fit of SK data}\label{atm-fit}
We now begin to study the SK data assuming, as usual,
that $\Delta m^2_{12}$ is too small to have relevant effects.
We start with a standard fit because, beyond being interesting,
allows to check our computation with other ones~\cite{SKexp,fogli,otherSKfits}.
We find a very satisfactory agreement with~\cite{fogli},
where many plots of oscillations effects are presented.

The oscillation parameters are precisely defined in the following way.
The neutrino mixing matrix $V$ is parametrized as
\begin{equation} V =
R_{23}(\theta_{23}) \pmatrix{1 &0&0 \cr 0&e^{i \phi}&0 \cr 0&0&1} 
R_{13}(\theta_{13})
R_{12}(\theta_{12})
\label{eq:V}
\end{equation}
where $R_{ij}(\theta_{ij})$ represents a rotation by $\theta_{ij}$ in the $ij$ plane
and $i=\{1,2,3\}$ are three neutrino mass eigenstates of mass $m_i$.
With this parameterization $\theta_{23}\sim45\degree$ gives the $\nu_\mu/\nu_\tau$ mixing tested at SK;
while $\theta_{1i}$ produce the deficit of solar $\nu_e$ neutrinos
($\theta_{13}$ could be zero).
We also define $\Delta m^2_{ij}\equiv m^2_j-m^2_i$.
We can assume that $|\Delta m^2_{23}|$ is the largest splitting (see~\cite{lungo} for more details).
With this parameterization atmospheric oscillations depend only on $\Delta m^2_{23}$,
$\theta_{23}$ and $\theta_{13}$.

\smallskip

In fig.~1 the continuous lines show how the $\chi^2$ depends on $\Delta m^2_{23}$
(for maximal $\theta_{23}=45\degree$ and zero $\theta_{13}$)
fitting separately the sub-GeV and the multi-GeV events.
Compared to a similar fit done by the SK collaboration,
our result is less sensitive to $\Delta m^2_{23}$ and allows slightly smaller values of $\Delta m^2_{23}$.
This small difference is probably due to the fact that
our fit does not include data about `upward through going muons'~\cite{mu-roccie}:
since our $\chi^2$ has a particularly flat minimum these less significant data,
that seem to prefer higher values of $\Delta m^2_{23}\sim 10^{-(3\div 1)}\eV^2$,
can cause the small difference.

The dotted lines in fig.~1 correspond to a simplified definition of the $\chi^2$:
we fit separately the angular dependence
of each one of the four kind of events measured by SK ($e$-like and $\mu$-like, sub-GeV and multi-GeV),
treating the overall normalization of each one of them as free
but including only statistical errors.
We see that, at least in the fit in fig.~1,
there is no significant difference between this simplified $\chi^2$ and the one in~\cite{fogli}.
An interesting feature shown by fig.~\ref{fig:chiq} is that {\em small values of
$\Delta m^2_{23}\circa{<}10^{-3}\eV^2$
(that are difficult to test at planned `long-baseline' experiments)
are now disfavoured by the clean multi-GeV data only\/}
(whose uncertainty is dominated by statistics),
rather than by the sub-GeV data
(whose interpretation strongly depends on the details of the SK experiment
and on theoretical predictions about the neutrino flux).
Thus the lower bound on $\Delta m^2$ is solid and
can be improved with more statistics in the next years.

\medskip

In fig.~\ref{fig:fitstandard} we show the results of a fit in
the relevant oscillations parameters,
$\Delta m^2_{23}$, $\theta_{23}$ and $\theta_{13}$.
The best fit has $\chi^2_{\rm min}=18$ (the $\chi^2$ uses 20 experimental data from SK and 2 from {\sc Chooz}).
We show contour lines corresponding to the values $\chi^2=24$ and $\chi^2=30$.
Using standard approximations, values of $\chi^2-\chi^2_{\rm min}$
can be converted into confidence levels that delimit `best fit regions', and values of $\chi^2$
can be converted into confidence levels that delimit `exclusion regions'.
Shaded areas roughly correspond to $(90\div 99)\%$ confidence levels\footnote{We do not insist on
the precise correspondence since it has no particular meaning.
There is no objective way of converting $\chi^2$ levels into statements like
``oscillation parameters lie in the shaded region with $90\%$ probability''.}.
In fig.~\hbox{\rm\ref{fig:fitstandard}a} we minimize the $\chi^2$ with respect to $\theta_{13}$
and determine the allowed regions in the plane $(\Delta m^2_{23},\theta_{23})$.
We see that $\sin^2 2\theta_{23}>0.8$ at $90\%$ C.L.
Fig.~\hbox{\rm\ref{fig:fitstandard}b} shows the $\chi^2$ minimized with respect to $\Delta m^2_{23}$
and determines the allowed values of the mixing angles.
In fig.~\ref{fig:fitstandard}c we minimize the $\chi^2$ with respect to $\theta_{23}$ and show
how the upper bound on $\theta_{13}$ depends on $\Delta m^2_{23}$.
If $\Delta m^2_{23}=3~10^{-3}\eV^2$ the {\sc Chooz} bound (dashed lines) requires a small value
of $\theta_{13}\circa{<}10\degree$.
For a mass splitting below the {\sc Chooz} bound, $\Delta m^2_{23}\circa{<}10^{-3}\eV^2$,
a larger $\theta_{13}$ is not forbidden by the {\sc Chooz} data,
but disfavoured by SK because it generates an up/down asymmetry in the $e$-like (sub-GeV and multi-GeV) sample.
However, as explained in the introduction,
the sub-GeV asymmetry vanishes for $\theta_{23}=45\degree$,
while the one in the multi-GeV sample vanishes for an appropriate value
of $\theta_{23}\circa{<}45\degree$.
Since our analysis does not say that $\Delta m^2_{23}=1\cdot 10^{-3}\eV^2$ is disfavoured,
for an appropriate value of $\theta_{23}\approx 45\degree$,
`large' $\theta_{13}\circa{>}20\degree$ are allowed.

\begin{figure*}[t]
\begin{center}
\begin{picture}(17.7,5)
\putps(-0.5,0)(-0.5,0){fit3masse}{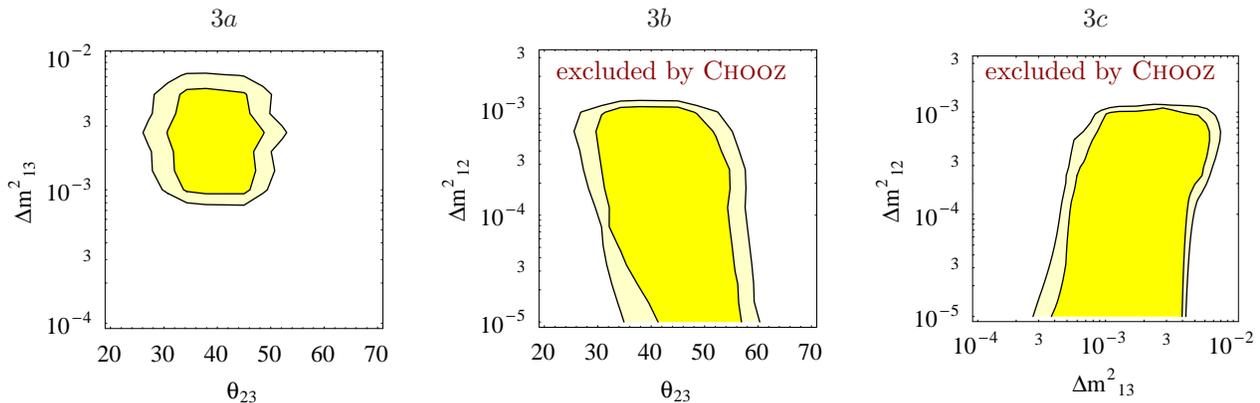}
\put(2.7,5.3){$\ref{fig:fit3masse}a$}
\put(8.5,5.3){$\ref{fig:fit3masse}b$}
\put(14.3,5.3){$\ref{fig:fit3masse}c$}\Red
\put(7.3,4.6){excluded by {\sc Chooz}}
\put(13,4.6){excluded by {\sc Chooz}}\Black
\end{picture}
\caption[SP]{\em Fit of SK data assuming that all $\Delta m^2$ are large enough to affect atmospheric neutrinos.
As in fig.~\ref{fig:fitstandard} we show contour lines of the $\chi^2(\Delta m^2_{ij},\theta_{ij})$ at $\chi^2=24$ and $30$
($\min\chi^2=17$).
All fits are restricted to $\theta_{13}=0$ and $\theta_{12}=45\degree$.
In fig.~\hbox{\rm\ref{fig:fit3masse}a} we show $\chi^2(\Delta m^2_{13},\theta_{23})$ for a
large value of $\Delta m^2_{12}=0.8~10^{-3}\eV^2$;
in fig.~\hbox{\rm\ref{fig:fit3masse}b (c)} we show the $\chi^2$ minimized with respect to $\Delta m^2_{13}$ ($\theta_{23}$).
\label{fig:fit3masse}}
\end{center}\end{figure*}

\begin{figure*}[t]
\begin{center}
\begin{picture}(17.7,5.8)
\putps(-0.2,0)(-0.2,0){f1245}{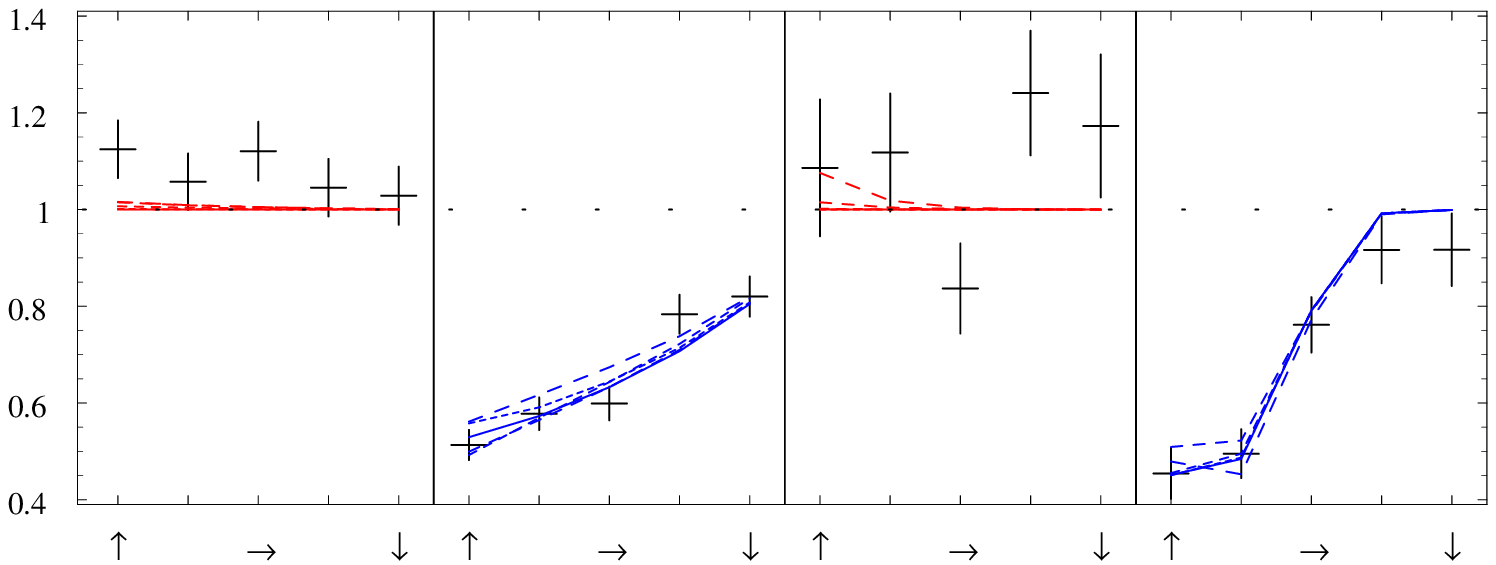}
\Red
\put(2.4,5.8){$e$-sub GeV}
\put(9.5,5.8){$e$-multi GeV}\Blue
\put(5.9,5.8){$\mu$-sub GeV}
\put(13.1,5.8){$\mu$-multi GeV}\Black
\end{picture}
\caption[SP]{\em Effect of `solar' oscillations on SK observables,
normalized to the unoscillated rates, for
$\theta_{12}=\theta_{23}=45\degree$, $\theta_{13}=0$,
$\Delta m^2_{13}=3~10^{-3} \eV^2$ and $\Delta m^2_{12}=\{0,0.3,1,3,9\}10^{-4}\eV^2$
(larger values have longer dashing).
The arrows on the horizontal axes denote the direction of incoming neutrinos.
The crosses are the experimental data (their error bars only include statistical errors).
\label{fig:sample}}
\end{center}
\begin{center}
\begin{picture}(17.7,5.8)
\putps(-0.2,0)(-0.2,0){f1230}{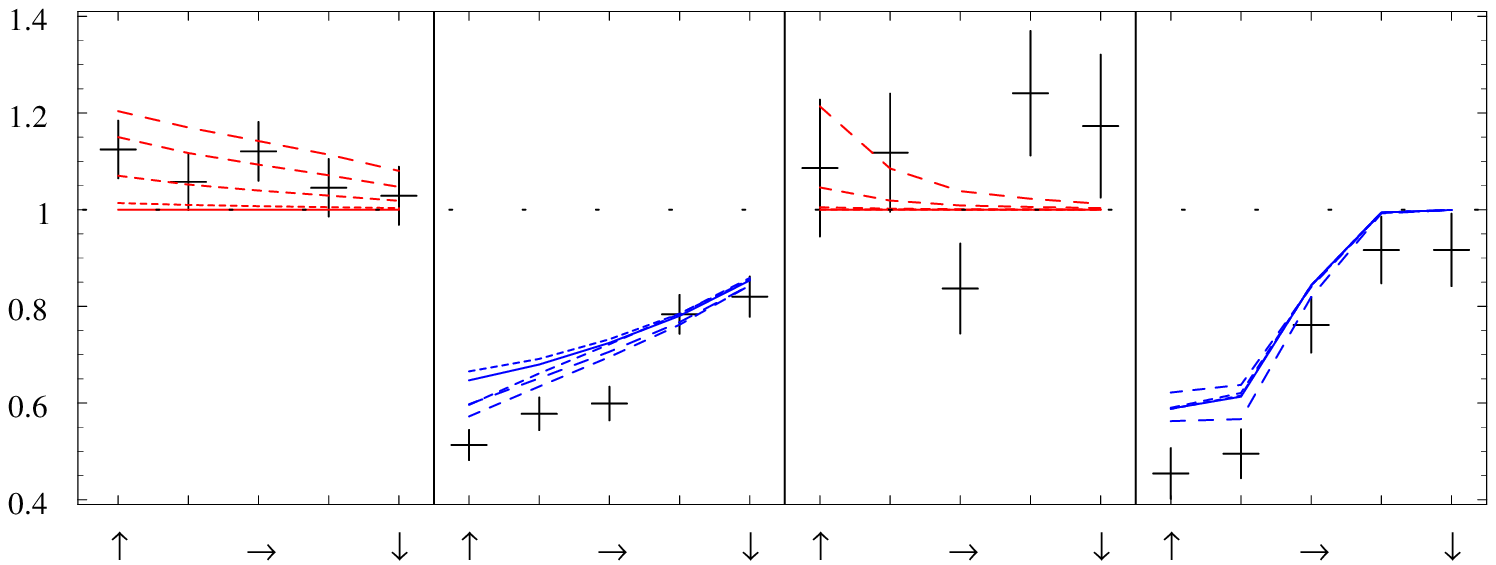}
\Red
\put(2.4,5.8){$e$-sub GeV}
\put(9.5,5.8){$e$-multi GeV}\Blue
\put(5.9,5.8){$\mu$-sub GeV}
\put(13.1,5.8){$\mu$-multi GeV}\Black
\put(4.8,4.2){\footnotesize$9$}
\put(4.8,3.95){\footnotesize$3$}
\put(4.8,3.7){\footnotesize$1$}
\put(4.75,3.45){\footnotesize$.3$}
\put(2,3.2){\footnotesize$\Delta m^2_{12}/(10^{-4}\eV^2)=0$}
\end{picture}
\caption[SP]{\em As in fig.~\ref{fig:sample}, but for a smaller $\theta_{23}=30\degree$.
The results for $\theta_{23}=60\degree$ are similar, but the sign of the effects is reversed.
\label{fig:sample2}}
\end{center}\end{figure*}

\section{Non-standard fit of SK data}\label{atm+sole}
Finally we assume that the `solar' $\Delta m^2_{12}$ is sufficiently large to
affect atmospheric neutrinos\footnote{Related analyses, motivated by the fact that the
large angle MSW solution allows $\Delta m^2_{12} < +0.2~10^{-3}\eV^2$
(the positive sign is the one that gives the desired MSW effect for solar neutrinos)
have been recently performed in~\cite{Giunti,Smirnov}.
For such values of $\Delta m^2_{12}$ our results agree with~\cite{Smirnov}
and disagree with~\cite{Giunti}.}.
The SK observables now depend on all oscillation parameters.
For simplicity we only exhibit fits
where $\theta_{12}=45\degree$
(as suggested by the deficit of solar neutrinos, if $\theta_{13}$ is small)
and $\theta_{13}=0$, so that the CP-violating phase $\phi$ becomes irrelevant.
In fig.s~\ref{fig:fit3masse} we show the result of our fit of SK atmospheric data
in $\Delta m^2_{13}$, $\Delta m^2_{12}$ and $\theta_{23}$.
Fig.~\ref{fig:fit3masse}a is done for
$\Delta m^2_{12}=0.8~10^{-3}\eV^2$, just below the {\sc Chooz} bound,
and shows that a good fit is possible for appropriate values of $\Delta m^2_{13}$ and $\theta_{23}$.
In fig.~\ref{fig:fit3masse}b (c) we minimize the $\chi^2$ with respect to $\Delta m^2_{13}$ ($\theta_{23}$).
The main result is that
{\em values of $\Delta m^2_{12}$ as large as $10^{-3}\eV^2$ are compatible
with the most recent SK data.}
Infact the upper bound on $\Delta m^2_{12}\circa{<} 0.9~10^{-3}\eV^2$ ---
shown in fig.s~\ref{fig:fit3masse}b,c --- comes from the {\sc Chooz} experiment.
Even for the maximal value of $\Delta m^2_{12}$,
`solar' oscillations have small effect on SK observables:
they produce an up/down asymmetry of $e$-like events
(this observable has a dominant statistical error) accompanied
by a change in the overall number of events
(this observable has a dominant theoretical error).
Depending on how the $\chi^2$ is defined this small effect can slightly improve or deteriorate the fit.

These small effects on $e$-like events produce the
preference for values of $\theta_{23}\circa{<}50\degree$ when $\Delta m^2_{23}$ is larger,
apparent from fig.s \ref{fig:fit3masse}a,b.
We now discuss this up/down asymmetry of $e$-like events in more detail because it is the most
promising signal of `solar' oscillations that can be observed at SK.

\medskip

The qualitative features of the asymmetry are well reproduced by the rough approximation
(again obtained making the simplifying assumptions used in~\cite{lungo})
\begin{equation}
{N_e^\uparrow\over N_e^\downarrow}\approx 1+\label{eq:Yesolar}
\frac{R(1+\cos2\theta_{23})-2}{4}\sin^22\theta_{12}.
\end{equation}
As already explained in the introduction, if $R=2$
(as in the sub-GeV data) and $\theta_{23}=45\degree$ there is no effect.
Multi-GeV neutrino events have $R\approx 3$; but
are rarer and too energetic for being strongly affected by a $\Delta m^2_{12}$ below the {\sc Chooz} bound.
Since the cancellation makes the situation intricate,
a numerical computation is necessary to determine
the possible signatures of a solar $\Delta m^2$ at SK.
We show in figure~\ref{fig:sample} some example of how the quantities measured by SK are affected
by `solar' oscillations for different values of $\Delta m^2_{12}$ if $\theta_{23}=45\degree$.
The arrows on the horizontal axes
denote the direction of the
incoming neutrinos and correspond to the five bins of $\cos\vartheta_{\rm zenith}$
used by the SK collaboration to present their results
(for example $\uparrow$ refers to the up-going neutrinos that cross the core of the earth).
The rate of each one of the 20 bins is normalized with respect to the no oscillation case
(our predictions for unoscillated rates are in satisfactory agreement with the Monte Carlo of the SK collaboration).
We see that the effect is very small for any value of $\Delta m^2_{12}$.

In figure~\ref{fig:sample2} we again show the effects of `solar' oscillations,
but for the smallest value of $\theta_{23}=30\degree$
compatible with the SK data.
There are now larger effects in the $e$ sub-GeV sample.
We see that a small $\sin^22\theta_{23}=3/4$ gives a poorer fit of $\mu$ events; but
this fit also depends on the value of $\Delta m^2_{23}$: for this reason
we do not consider useful discussing small `solar' effects in the $\mu$ events.
To correctly interpret fig.~\ref{fig:sample2}, we must remind that the overall normalization of the fluxes
(i.e.\ the `1' line in the plot) has a $\sim20\%$ theoretical uncertainty.
Moreover the `1' lines for the four different data samples can be moved independently by $\sim5\%$.
Our $\chi^2$ knows that these systematic uncertainties are highly correlated
and says that (with present statistics) even the `solar' effects shown in fig.~\ref{fig:sample2} are ``small effects''.
For $\theta_{23}=60\degree$ the effects due to `solar' oscillations have similar size, but opposite sign.

\medskip

In all these computation we have assumed that $\Delta m_{23}^2,\Delta m^2_{12}>0$.
Matter effects do not significantly affect the results if the $\Delta m^2$ have different signs.

\begin{figure*}[t]
\begin{center}
\begin{picture}(17.7,5)
\putps(-0.5,0)(-0.5,0){f1340}{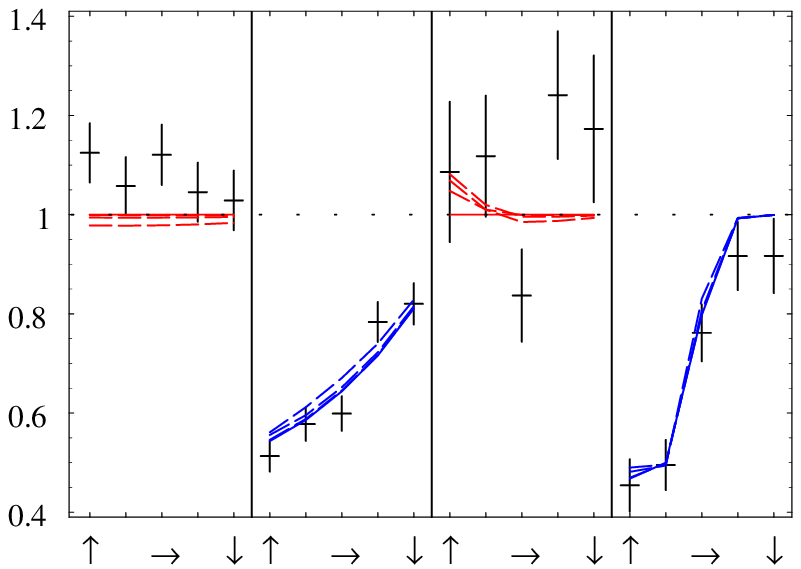}
\putps(8.5,0)(8.5,0){f1350}{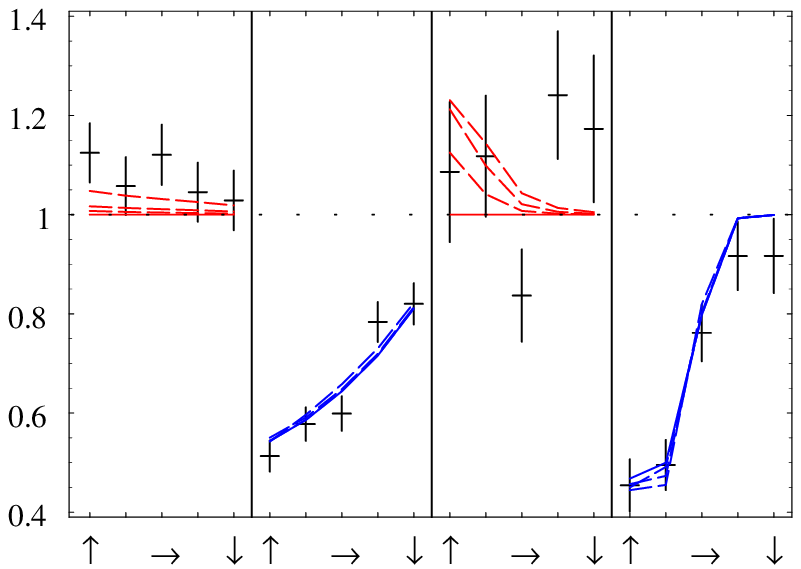}
\put(1.8,5.9){$\theta_{23}=40\degree$, $\theta_{13}=\{0,5\degree,10\degree,20\degree\}$}
\put(10.5,5.9){$\theta_{23}=50\degree$, $\theta_{13}=\{0,5\degree,10\degree,20\degree\}$}
\Red
\put(0.5,1){$e$-subGeV}
\put(9.5,1){$e$-subGeV}
\put(4.1,1){$e$-multiGeV}
\put(13.1,1){$e$-multiGeV}\Blue
\put(2.4,5.2){$\mu$-subGeV}
\put(5.9,5.2){$\mu$-multiGeV}
\put(11.3,5.2){$\mu$-subGeV}
\put(14.9,5.2){$\mu$-multiGeV}\Black
\end{picture}
\caption[SP]{\em Sample of how the SK observables are affected by
a non zero $\theta_{13}$ for $\Delta m^2_{23}=3~10^{-3} \eV^2$
and $\theta_{23}=40\degree$ (left)
$\theta_{23}=50\degree$ (right).
The arrows on the horizontal axes denote the direction of the incoming neutrinos.
\label{fig:sample13}}
\end{center}\end{figure*}

\subsection{Non zero $\theta_{13}$}
So far we have assumed that $\theta_{13},\phi=0$.
Like in the `standard' scenario,
a small $\theta_{13}\circa{<}20\degree$ is allowed by the {\sc Chooz} and SK data.
CP violation cannot affect SK observables since integration over neutrino energy averages it to zero.

We now discuss how effects due to $\theta_{13}$ mixing 
can be distinguished from effects due to a `solar' oscillation.
In fig.~\ref{fig:sample13} we show how a $\theta_{13}=\{0,5\degree,10\degree,20\degree\}$ affects
the SK observables if $\Delta m^2_{12}\ll\Delta m^2_{23}=3\*10^{-3}\eV^2$, $\theta_{23}=\{40\degree,50\degree\}$.
Comparing the results of the numerical computation
shown in fig.~\ref{fig:sample13} with 
fig.s~\ref{fig:sample},\ref{fig:sample2}, we notice two main differences
\begin{itemize}

\item $\theta_{13}$ oscillations mainly affect multi-GeV $e$-like events,
while `solar' oscillations can only affect $e$ sub-GeV events.

\item For a given value of $\theta_{23}$, `solar oscillations' and `$\theta_{13}$-oscillations'
produce up/down asymmetries of opposite sign.
The up/down asymmetry produced by $\theta_{13}$ is different from zero and positive even if $\theta_{23}=45\degree$.

\end{itemize}
(In particular, a not too large $\theta_{13}$ can only produce
a few $\%$ excess of sub-GeV $e$-like events).
Both these features can be understood comparing the approximate up/down asymmetries produced by
a small $\theta_{13}$,
\begin{equation}
{N_e^\uparrow\over N_e^\downarrow}\approx 1+2\theta_{13}^2(R\sin^2\theta_{23}-1),\label{eq:Ye13}
\end{equation}
with the corresponding approximation for `solar' effects, eq.\eq{Yesolar}.
We remind that $\Delta m^2_{23}\circa{>} \Delta m^2_{12}$
and that sub-GeV events have $R\approx 2$, while multi-GeV ones have $R\approx 3$.
In both cases the effects
cancel out if $R\approx 2$ and $\theta_{23}\approx 45\degree$.

\section{Conclusions}
In conclusion a large `solar' $\Delta m^2_{12}\circa{<} 10^{-3}\eV^2$
is not safely excluded by solar neutrino experiments and is allowed by atmospheric neutrino experiments\footnote{On
the contrary no acceptable fit of SK data is possible for the mass pattern
$\Delta m^2_{12}\sim10^{-3}\eV^2$ and $\Delta m^2_{23}\sim \eV^2$,
sometimes invoked for reconciling the unconfirmed LSND oscillation with
solar and atmospheric ones.
We do not show any numerical result
because this fact is sufficiently clear from the approximate analysis in~\cite{lungo}.}
A `solar' oscillation has little effect on SK observables and can slightly ameliorate or deteriorate the fit of SK data,
depending on how the $\chi^2$ is defined.
Its most clear signature at SK is an up to $\sim15\%$ angular dependent excess (or deficit)
of $e$ sub-GeV events.
An indication for a  $\sim10\%$ excess of $e$-like events in the sub-GeV sample
was present in the first year of data taking at SK;
but the evidence has decreased in the most recent analyses with doubled statistics.
However this happened in a not very nice way:
\begin{itemize}

\item The SK collaboration has introduced small improvements in their Monte carlo,
and obtained slightly different predictions;

\item The rate of $e$ sub-GeV events in the most recent part of the sample
(last 321 days of data taking) is $(18\pm 5)\%$ lower than in the first part (first 414 days)~\cite{LoSecco}.

\end{itemize}
Moreover, if the overall flux of atmospheric $\nu$ is somewhat lower than what current estimates indicate ---
as suggested by a recent (preliminary) measurement of cosmic ray fluxes~\cite{BESS} ---
than SK is observing a smaller deficit of $\nu_\mu$ and an excess of $\nu_e$ events.
In conclusion we believe that the normalization of $\nu$ fluxes is still very uncertain and that
an excess (or even a deficit) of $e$ sub-GeV events could be present.
With more statistics it will be possible to use the unoscillated down-going multi-GeV events to fix the normalization,
or to search for up/down asymmetries in the $e$ samples.

\medskip

\begin{figure*}[t]
\begin{center}
\begin{picture}(17.7,4)
\putps(-0.5,0)(-0.5,0){fMinos}{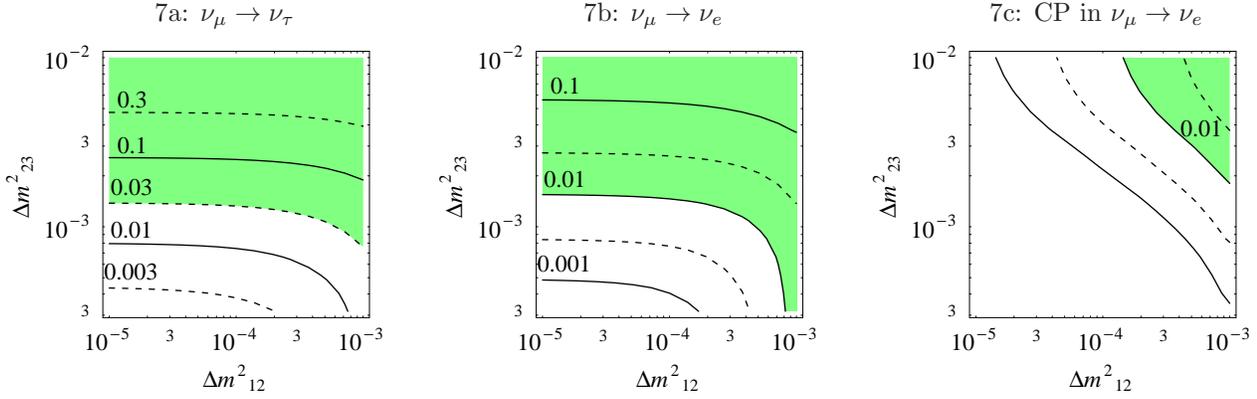}
\put(2,5.3){\ref{fig:fMinos}a: $\nu_\mu\to\nu_\tau$}
\put(7.7,5.3){\ref{fig:fMinos}b: $\nu_\mu\to\nu_e$}
\put(13.1,5.3){\ref{fig:fMinos}c: CP in $\nu_\mu\to\nu_e$}\Red
\Black
\end{picture}
\caption[SP]{\em Contour plots of the transition probabilities $P(\nu_\mu\to \nu_\tau)$
\hbox{(fig.~\rm\ref{fig:fMinos}a)}, CP-conserving part of $P(\nu_\mu\to \nu_e)$ \hbox{(fig.~\rm\ref{fig:fMinos}b)} and
CP-violating part of $P(\nu_\mu\to \nu_e)$ \hbox{(fig.~\rm\ref{fig:fMinos}c)} at {\sc Minos} (`medium' beam)
as function of the mass splittings $(\Delta m^2_{12},\Delta m^2_{23})$ for
maximal CP-violatng phase $\phi=90\degree$,
maximal $\theta_{12}=\theta_{23}=\pi/4$ and moderately large $\theta_{13}=20\degree$.
For this value of $\theta_{13}$, $\Delta m^2_{23}\circa{>}3~10^{-3}$ is disfavoured by {\sc Chooz}.
{\sc Minos} can explore the shaded regions.
\label{fig:fMinos}}
\end{center}\end{figure*}

To conclude, a large `solar' $\Delta m^2_{12}$ can be experimentally investigated in different ways:
\begin{itemize}
\item {\bf SuperKamiokande},  when high-statistics will be available,
could see some indication in the details of $e$ events;

\item future {\bf solar experiments} (like Borexino) could exclude this possibility;

\item {\bf KamLand} could soon observe an evident deficit of reactor $\bar{\nu}_e$;

\item `{\bf long-baseline experiments}' (like {\sc K2K} and {\sc Minos}) can confirm the
$\nu_\mu \leftrightarrow\nu_\tau$ oscillation seen at SK.
These experiments could also see the $\nu_e$ appearance
signal due to a sufficiently large $\theta_{13}$.
On the contrary $\nu_e$ appearance\footnote{The $\nu_\mu$ disappearence observed at SK reduces
the $\nu_e$ appearance signals at {\sc K2K} and {\sc Minos} by a factor close to $2$.
This explains why, according to fig~\ref{fig:fMinos}b,
these experiments are less sensitive to $\nu_\mu\to\nu_e$ oscillations than in
standard estimates, done using the `two-neutrino approximation' where $\nu_\tau$ are neglected.}
due only to `solar' effects
can be seen only if $\Delta m^2_{12}$
is very close to its maximal value, $10^{-3}\eV^2$.
CP violating effects in $\nu_\mu\to \nu_e$ are too small to affect the experiment,
except maybe in some extreme case~\cite{rom}.
All these conclusions are illustrated by fig.~\ref{fig:fMinos}, where we show contour-plots of the
relevant transition probabilities averaged over
the energy spectrum of the `medium energy' $\nu_\mu$ beam at {\sc Minos}
(see the caption for more details).

\item future `{\bf neutrino factories}'
(beam of $\bar{\nu}_\mu\nu_e$ produced by decay of muons)~\cite{nu-factory},
if $\Delta m^2_{12}$ is large enough,
could study CP violation and make precision measurements of the oscillation parameters
(but it could be difficult to distinguish $\theta_{13}$ effects from $\Delta m^2_{12}$ effects).

\end{itemize}

\paragraph{Note added}
In the recent paper~\cite{3maxmix} it was observed that if $\Delta m^2_{23}\approx 10^{-3}\eV^2$
the (interesting?) scenario named `tri-maximal mixing'
(in our language $\theta_{12}=\theta_{23}=\pi/4$ and $\sin^2\theta_{13}=1/3$, i.e.\
$\theta_{13}\approx 35\degree$) is still compatible with SK and {\sc Chooz} data
(at least until the complete {\sc Chooz} data will be presented),
due to certain cancellations.
This is also shown by fig.~\ref{fig:fitstandard}c:
as we have discussed, these cancellations are not a special feature of tri-maximal mixing.
Quite generally SK is rather insensitive to new oscillations beyond the observed one.

\paragraph{Acknowledgments}
I thank R. Barbieri, D. Nicol\`o and A. Yu Smirnov  
for clarifying discussions.

\end{document}